\documentclass[12pt]{article}
\pdfoutput=1
\usepackage{yfonts}
\usepackage{color}
\usepackage{mhchem}
\usepackage{xcolor}
\usepackage{mathrsfs}
\usepackage[mathscr]{eucal}
\usepackage{cite}
\usepackage{hyperref}
\hypersetup{colorlinks=true,linkcolor=red,anchorcolor=black,citecolor=green}
\usepackage[toc,page]{appendix}
\usepackage{amsfonts}
\usepackage{bbold}
\usepackage{textcomp}
\usepackage[DIV13]{typearea}
\usepackage{amsmath, amsthm, amssymb, mathtools,empheq,latexsym,dsfont}
\usepackage{bbm}
\usepackage{slashed, simplewick}
\usepackage[utf8]{inputenc}
\usepackage{graphicx,placeins}
\usepackage{makeidx}
\usepackage[font=small,labelfont=bf]{caption}
\usepackage{nicefrac}
\usepackage{subfigure}
\usepackage{array, bigdelim,multirow,multicol}
\usepackage[integrals]{wasysym}
\usepackage{fancybox}
\usepackage{bm}
\usepackage{float}
\usepackage{rotating}
\usepackage{colortbl}
\usepackage{booktabs}
\usepackage[top=2cm,textwidth=16.6cm,textheight=22.75cm]{geometry}
\usepackage{doi}
\graphicspath{{immagini/}}

\definecolor{Gray}{gray}{0.92}
\newcommand{\nc}{\newcommand}
\nc{\Dfb}{\mbox{$\raisebox{2mm}{\boldmath ${}^\leftrightarrow$}\hspace{-4mm} D$}}

\def\dd{\displaystyle}
\nc{\btb}{\begin{tabular}}    \nc{\etb}{\end{tabular}}

\newcommand{\nn}{\nonumber}
\newcommand{\be}{\begin{equation}}
\newcommand{\ee}{\end{equation}}
\newcommand{\bea}{\begin{eqnarray}}
\newcommand{\eea}{\end{eqnarray}}

\makeatletter
\renewcommand*{\@fnsymbol}[1]{\ensuremath{\ifcase#1\or *\or  \mathsection\or \ddagger\or
\dagger\or \mathparagraph\or \|\or **\or \dagger\dagger
\or \ddagger\ddagger \else\@ctrerr\fi}}
\makeatother

\makeatletter
\@addtoreset{equation}{section}
\makeatother

\begin{document}
 \unitlength = 1mm

\setlength{\extrarowheight}{0.2 cm}

\title{
\begin{flushright}
\begin{minipage}{0.2\linewidth}
\normalsize
\end{minipage}
\end{flushright}
 {\Large\bf A Note on Gauge Anomaly Cancellation\\[0.3cm]
in Effective Field Theories}\\[1.3cm]}
\date{}

\author{
Ferruccio~Feruglio$^{1}$
\thanks{E-mail: {\tt feruglio@pd.infn.it}}
\\*[20pt]
\centerline{
\begin{minipage}{\linewidth}
\begin{center}
$^1${\small
Dipartimento di Fisica e Astronomia `G.~Galilei', Universit\`a di Padova\\
INFN, Sezione di Padova, Via Marzolo~8, I-35131 Padua, Italy}\\*[10pt]
\end{center}
\end{minipage}}
\\[10mm]}
\maketitle
\thispagestyle{empty}

\centerline{\large\bf Abstract}
\begin{quote}
\indent
The conditions for the absence of gauge anomalies in effective field theories (EFT) are rivisited.
General results from the cohomology of the BRST operator do not prevent potential
anomalies arising from the non-renormalizable sector, when the gauge group is not semi-simple, like in the Standard Model EFT (SMEFT).
By considering a simple explicit model that mimics the SMEFT properties, we compute the anomaly in the regularized theory,
including a complete set of dimension six operators.
We show that the dependence of the anomaly on the non-renormalizable part can be removed by adding a local counterterm to the theory.
As a result the condition for gauge anomaly cancellation is completely controlled by the charge assignment of the
fermion sector, as in the renormalizable theory.
\end{quote}

\newpage

\section{Introduction}
In the last few years, the role of EFT in the description of particle interactions has become even more relevant than it has been in the past, when specific scenarios like
low-energy supersymmetry or composite dynamics in the electroweak symmetry breaking sector were dominating
the searches for New Physics (NP). Within few assumptions concerning symmetries, field content and power counting, EFT allow to
parametrize possible NP effects in the most general way. Any theory of this type includes potentially
an infinite number of parameters $c_k$, related to the independent operators of increasing dimensionality consistent
with the assumed symmetries and field content. 
Although information on these parameters from the
experimental side is clearly crucial, 
it is also very important to understand whether theoretical bounds apply to them. Among such bounds,
of primary importance are those related to fundamental properties of any respectable quantum field theory, such as causality, unitarity and, in the particle physics context, Lorentz invariance.  Indeed, recently there has been a great activity in deriving limits
on $c_k$ related to analyticity, unitarity and crossing properties of scattering amplitudes
\cite{Adams:2006sv,Vecchi:2007na,Nicolis:2009qm,Bellazzini:2016xrt,deRham:2017avq,deRham:2017zjm,Zhang:2018shp,Bi:2019phv,Bellazzini:2020cot,Remmen:2019cyz,Remmen:2020uze,Zhang:2020jyn}. These bounds typically apply to coefficients of operators of dimension greater or equal to eight and
define a permitted region, or landscape, in theory space. 

Another fundamental requirement of the present description of particle interactions is gauge invariance.
Strong and electroweak interactions are described by theories invariant under local continuous transformations.
Gauge invariance can be realized in the exact or in the spontaneously broken phase, but in either cases it
should be free from anomalies. Anomalies arise from fermion loops \cite{Adler:1969gk,Bell:1969ts}. In renormalizable theories there is a well-known criterium for the absence of gauge
anomalies \cite{Georgi:1972bb}. It entirely relies on the transformation properties (charges and generators) of fermion fields,  
and it is independent from the parameters characterizing the theory. When moving from the renormalizable case
to the non-rinormalizable one, comprising the whole set of EFTs, we might ask whether this criterium
is still sufficient to ensure the absence of gauge anomalies, or whether it should be complemented by an additional set of 
conditions on the coefficients $c_k$. If the latter possibility applied, we would have discovered new bounds,
sharping the landscape of admissible theories. In particular all this discussion applies to the SMEFT,
which currently provides one of the most reliable tools to parametrize NP. 

Given the relevance of such a question, it is not surprising that it has been discussed at length in the literature, also 
in relation with the SMEFT.  Anomalies are non-trivial BRST cohomology classes of ghost number one, in the space of local functionals of the theory. The BRST cohomology for Yang-Mills and gravity theories coupled to scalars and fermions is known for all dimensions of space-time and for all possible local polynomial interactions, thus including the full set of possible EFT. The result is that for gravity and for semi-simple non-abelian gauge theories, all possible Lorentz-invariant anomalies are polynomials of dimension equal to the dimension of space-time \cite{Barnich:1994ve,Barnich:2000zw}. There are only a finite number of them for a given theory, and they are all well known. As we will argue, this strongly disfavor the possibility that gauge anomalies, for semi-simple groups, depend on the coefficients $c_k$ in questions. A very different result holds when there is an abelian factor in the gauge group. Then there are potential anomalies of all dimensions \cite{Dixon:1979zp} and the dependence on $c_k$ requires further investigation, beyond cohomological
arguments \cite{Dixon:1991wu}. This in particular applies to the SMEFT, whose gauge group is not semi-simple.

The authors of ref. \cite{QED} computed the axial anomaly in a non-renormalizable version of quantum electrodynamics (QED), finding the same result as in renormalizable QED. 
Chiral anomalies in theories involving higher-derivative couplings with non-abelian gauge fields have been studied in ref. \cite{Kim:1988xy}. The corresponding expressions for the covariant and consistent anomalies agree with those based on the minimal Lagrangian.
An extensive discussion of anomalies arising in non-renormalizable theories can be found in ref. \cite{Minn:1986ba}. The authors show that in 2$d$-dimensional spacetime the inclusion of a general set of local, gauge invariant, non-renormalizable operators does not lead to any new gauge anomaly, besides the ones already identified from the renormalizable part. They conclude that there is no restriction on
the parameters of these operators from the requirement of gauge anomaly cancellation. Two arguments supporting this result are presented in ref. \cite{Minn:1986ba}. One comes from the analysis of the imaginary part of the Euclidean effective action \cite{AlvarezGaume:1983ig}. The other one relies on the discussion of triangle diagrams and exploits two different
regularization procedures.  Diagrams involving only renormalizable couplings are dealt with a regulator that breaks gauge invariance, while diagrams including some non-renormalizable interaction are discussed within a gauge-invariant regularizing
framework. It would be interesting to see what is the outcome when a unique regularization is used for all diagrams. This is actually one of the points illustrated in the present note.
 In ref. \cite{Soto:1990ij} anomalies arising from truncating the SMEFT at dimension six are discussed. No genuine contribution to the anomaly arises from dimension six operators. Anomaly cancellation at this order does not relate coupling constants of dimension six operators as one might naively expect, and hence does not constrain the physics beyond the SM. Nevertheless, the argument of ref. \cite{Soto:1990ij} cannot be considered conclusive, since the operators
analyzed in this work, carrying no dependence on the Higgs field, do not form a complete set.

Dimension six operators depending on scalar fields, such as $c_k(\varphi^\dagger \cdot D_\mu\varphi-
D_\mu\varphi^\dagger\cdot\varphi)\bar\psi_k \gamma^\mu\psi_k$, have been recently discussed in ref. \cite{Cata:2020crs}.
The authors observed that, in the spontaneously broken phase, the presence of these operators leads to a shift
of the fermion gauge couplings proportional to the parameters $c_k$. On this basis they argue that 
the cancellation of gauge anomalies requires new conditions involving $c_k$.
A very recent paper \cite{Bonnefoy:2020tyv} shows that these conditions are violated in a number of consistent, anomaly-free, models of NP. The authors of ref. \cite{Bonnefoy:2020tyv} also show that the premature conclusion of ref. \cite{Cata:2020crs} derives from the non-inclusion of the (would-be) Goldstone bosons in the expression of the gauge current. 
This analysis, which goes in the right direction, is however still incomplete, since it does not include a general background
in the computation of the anomaly. Indeed, as we shall see in detail here, the full bosonic background can reintroduce the potential dependence of the anomaly on the parameters $c_k$. This is an important point,
perhaps the main one we would like to clarify here.

Purpose of this work is to evaluate the gauge anomaly in an EFT including a full set of dimension six-operators and allowing
for the most general bosonic background, a task not yet carried out in the literature.
This will be done in a simple model with U(1) gauge invariance displaying all the features of the SMEFT:
i) gauge invariance realized in the broken or unbroken phase, depending on the region in parameter space; ii) a set of chiral fermions including ''quarks'' and "leptons"; iii) cancellation of gauge anomalies in the renormalizable part requiring a specific assignment of U(1) fermion charges. We discuss the properties of the regularized effective action, obtained by integrating
over the fermionic degrees of freedom. By an explicit one-loop computation, carried out within dimensional regularization, we derive the variation of this functional under a gauge transformation. The result depends on the full set of bosonic fields, scalars and gauge boson.
As we shall see, even using the correct form of the current advocated in ref. \cite{Bonnefoy:2020tyv}, there are new contributions depending on $c_k$ and we can still wonder whether their cancellation requires
a condition on $c_k$. The main result of the present work is to show how all the contributions depending on the coefficients of the higher-dimensional operators are trivial, since they can be canceled by adding to the regularized effective action a local polynomial in the bosonic fields. The cancellation of the remainder, non-trivial, part requires the well-known condition between the fermion charges. No other relations between Lagrangian parameters are needed.
\section{A heuristic argument}
Given a relativistic quantum field theory, renormalizable or not, invariant under the action of a gauge group $G$ and depending on a set of
fields $(A_\mu,\varphi,\psi)$ describing particles of spin (1,0,1/2), the gauge invariance of the related classical
action $S$ can be expressed through a set of local operators $L(x)$~\footnote{Indices are omitted, when not essential.}:
\be
\delta_\alpha S=\int d^4x~ \alpha(x)L(x)S=0~~~.
\label{classical}
\ee
The classical, covariantly conserved, gauge currents are defined as:
\be
j^\mu(x)=-\frac{1}{g}\frac{\delta S}{\delta A_\mu(x)}~~~.
\ee
In the quantized theory, anomalous contributions are caused by fermion loops and
can be studied by means of the effective action $W[A_\mu,\varphi,\varphi^\dagger]$ defined through the path integral:
\be
e^{\dd i W[A_\mu,\varphi,\varphi^\dagger]}=\int {\cal D}\psi {\cal D}\bar\psi e^{\dd iS}~~~,
\label{effact}
\ee
where the integrated fields are the fermionic ones. If we are interested in the divergence
of the gauge current in a general background $B=(A_\mu,\varphi,\varphi^\dagger)$ at the lowest non-trivial order, we do not need to integrate over the bosonic fields. 
In particular we do not address here possible contributions to the anomaly from higher loops \cite{Adler:1969er,Anselmi:2015sqa}
and we do not need to quantize the fields $(A_\mu,\varphi,\varphi^\dagger)$. We will comment on
this point later on. Gauge anomalies express the non-invariance of the effective action $W[A_\mu,\varphi,\varphi^\dagger]$ under gauge transformations:
\be
\delta_\alpha W=\int d^4x~ \alpha(x)L(x)W[A_\mu,\varphi,\varphi^\dagger]\ne 0~~~.
\label{ward0}
\ee
Care should be taken when computing the gauge variation $L(x)W[A_\mu,\varphi,\varphi^\dagger]$. This expression is
formally divergent and requires a regularization. First, what we really compute is rather:
\be
L(x)W_{\tt r}[A_\mu,\varphi,\varphi^\dagger]~~~,
\label{ward0}
\ee
where $W_{\tt r}$ is the regularized version of $W$~\footnote{A limiting procedure where the regulator is removed after
the evaluation of $L(x)W_{\tt r}[A_\mu,\varphi,\varphi^\dagger]$ is understood here.}. The renormalized effective action $W[A_\mu,\varphi,\varphi^\dagger]$ and its gauge variation are recovered by adding to $W_{\tt r}$ the space-time integral of local polynomials in the fields $(A_\mu,\varphi,\varphi^\dagger)$, thus fixing the renormalization scheme. If we can find a local polynomial $P_{\tt c}$ such that
$L(x)(W_{\tt r}+\int d^4y P_{\tt c}(y))=0$, we can define
$W=W_{\tt r}+\int d^4y P_{\tt c}(y)$ and the theory is free from gauge anomalies. In this case, $L(x)W_{\tt r}$ is also said an irrelevant anomaly. Thus relevant anomalies
are non-trivial classes $\{L(x)W_{\tt r}\}$ under the equivalence $L(x)W_{\tt r}\sim L(x)W_{\tt r}'$ where $W_{\tt r}'=W_{\tt r}+\int d^4y P(y)$, $P(y)$ being a local polynomial in the bosonic fields. In general, we expect $L(x)W_{\tt r}$ to contain relevant and irrelevant contributions. 

The functional dependence of $L(x)W_{\tt r}$, is strongly constrained by the Wess-Zumino consistency conditions \cite{Wess:1971yu}:
\be
L_a(x)L_b(y)W_{\tt r}-L_b(y)L_a(x)W_{\tt r}= \delta^4(x-y)f_{ab}^cL_c(x)W_{\tt r}~~~,
\label{WZ}
\ee 
consequence of the algebra of the operators $L(x)$. Here $f_{ab}^c$ are the structure constants of the gauge group. We can regard the class $\{L_a(x)W_{\tt r}\}$ as the unknown in eq.
(\ref{WZ}). The general solution can be derived from cohomological arguments~\footnote{In the fully quantized theory, gauge invariance is replaced by BRST invariance, and eq. (\ref{WZ}) is replaced by $\delta_{BRST}^2 W_{\tt r}=0$,
$\delta_{BRST}$ being the nilpotent BRST operator.}. 
For semi-simple non-abelian gauge theories, renormalizable or not, all possible Lorentz-invariant anomalies are polynomials of dimension equal to the dimension of space-time \cite{Barnich:2000zw}. They coincide with the well-known Adler-Bell-Jackiw anomalies \cite{Barnich:1994ve}. Moreover, by the non-renormalization theorem of Adler-Bardeen \cite{Adler:1969er} and its generalization to non-renormalizable theories \cite{Anselmi:2015sqa}, such anomalies are exhausted by one-loop contributions. These results are the ingredients of a heuristic argument excluding the dependence
of the anomaly on the coefficients $c_k$ controlling non-renormalizable operators, for semi-simple gauge groups.

The argument goes as follows. In a general EFT the relevant fermion bilinear interaction reads:
\be
\bar f \left(\mathscr{S}+\mathscr{P} \gamma_5+\mathscr{V}_\mu \gamma^\mu+\mathscr{A}_\mu \gamma^\mu\gamma_5+\mathscr{T}_{\mu\nu} \sigma^{\mu\nu} \right)f~~~,
\ee
where $\mathscr{S}$, $\mathscr{P}$, $\mathscr{V}_\mu$, $\mathscr{A}_\mu$, $\mathscr{T}_{\mu\nu}$
are polynomials in the bosonic fields and their derivatives. Only the $\mathscr{V}_\mu$, $\mathscr{A}_\mu$ couplings matter for the anomaly. We can expand them in contributions coming from operators of increasing dimensionality:
\bea
\mathscr{V}_\mu&=&c_0^V A_\mu+\Omega_\mu^V~~~~~~~~~~\Omega_\mu^V=\sum_{k>0} c_k^V {\cal O}_{k,\mu}^V\nn\\
\mathscr{A}_\mu&=&c_0^A A_\mu+\Omega_\mu^A~~~~~~~~~~\Omega_\mu^A=\sum_{k>0} c_k^A {\cal O}_{k,\mu}^A~~~.
\eea
Here the coefficients $c_0^{V,A}$ come from the renormalizable part and have zero mass dimension. They are fixed by the minimal coupling
between fermions and gauge bosons, completely determined by the gauge transformation properties of the fermion fields.
The parameters $c_k^{V,A}$ $(k>0)$ derive from the non-renormalizable sector and have mass dimension $-k$.
The operators ${\cal O}_{k,\mu}^{V,A}$ have dimension $k+1$. The functional $W_{\tt r}$, evaluated at one-loop order, can be expanded in powers of $\mathscr{V}_\mu$ and $\mathscr{A}_\mu$. The gauge variation of any given order of such an expansion can be decomposed into two parts. The first one
depends only on the gauge field $A_\mu$. If not vanishing, this part should be a polynomial of degree four in $A_\mu$
and its derivatives. The second one originates also from $\Omega_\mu^{V,A}$ and will contain some monomial in fields and derivatives of degree higher than four,
contrary to the general cohomological results. Therefore either these new monomials are vanishing or they are irrelevant, and we conclude that there is no dependence on $c_k^{V,A}$ $(k>0)$ in the anomaly.

In this work we will not rely on the previous heuristic argument, since in any case it does not apply to non semi-simple
gauge groups like the one of the SMEFT. The solutions of the Wess-Zumino consistency conditions
in the abelian case allow potential anomalies of any dimensionality. Indeed, the right-hand side of eq. (\ref{WZ})
vanishes for abelian groups and $L(x) W_{\tt r}$ can be any gauge invariant polynomial in $A_\mu$, $\varphi$,
$\varphi^\dagger$, whose dimension is not bounded. It is not difficult to build some candidates for relevant anomalies $\{L(x) W_{\tt r}\}$.
Consider a U(1) gauge theory where $\varphi$ is a complex scalar field carrying a non-vanishing charge.
The following local operators:
\be
\varphi^\dagger\varphi~ \varepsilon^{\mu\nu\rho\sigma}\partial_\mu A_\nu~\partial_\rho A_\sigma~~~,~~~~~~
i~\varphi^\dagger\varphi~ \varepsilon^{\mu\nu\rho\sigma}\partial_\mu (\varphi^\dagger D_\nu\varphi- D_\nu\varphi^\dagger \varphi) ~\partial_\rho A_\sigma~~~,
\label{cand}
\ee
are solutions of eq. (\ref{WZ}) and cannot be expressed as gauge variations of an integrated local polynomial. The previous heuristic argument does not apply to these examples. These expressions can only come from 
the contribution of higher-dimensional operators to the anomaly and, if present, carry a dependence on the
coefficients $c_k$. Their occurrence in a given model can only be verified through a direct computation, which constitutes the main aim
of this paper. By considering a simple explicit model that mimics the SMEFT properties, we will see that $L(x)W_{\tt r}$
is the sum of two contributions. The first one includes only the gauge fields $A_\mu$ and is independent
from the Lagrangian coefficients $c_k$. It leads to the well-known conditions for gauge anomaly cancellation.
The second one involves both $A_\mu$ and $\varphi$ and depends on the coefficients $c_k$. We explicitly
show that this part is irrelevant and does not require additional conditions for the absence of gauge anomalies.
A posteriori, this provides a confirmation of the above heuristic argument, at least in a specific case.

\section{A miniature SMEFT}
To illustrate the mechanism of anomaly cancellations in EFT, we consider a toy model displaying many features
of the SMEFT, but sufficiently simple to allow a concise description of the problem. 
The model enjoys a U(1)$_Q$ gauge symmetry. 
The matter fields consist of
two four-component fermions, $l$ and $q$, mimicking leptons and quarks in the SMEFT, plus a complex scalar $\varphi$,
the analogue of the SMEFT Higgs multiplet. In a four-component notation, $Q(l_L)=-Q(q_L)=Q(\varphi)=-1$
and $Q(l_R)=Q(q_R)=0$. The gauge theory is chiral but with this assignment gauge anomalies generated by the renormalizable part of the theory are absent.
In what follows we will focus on the sector of the theory where baryon and lepton numbers $B$ and $L$ are conserved~\footnote{At variance with the SM, the conservation of $B$ and $L$ at the renormalizable level are not automatic in our setting, but it could be
easily enforced by suitable discrete symmetries.}.
The effective Lagrangian reads:
\be
{\cal L}={\cal L}_4+{\cal L}_6+...
\ee
where ${\cal L}_4$ denotes the renormalizable part, ${\cal L}_6$ collects dimension six
operators and dots stand for higher-dimensional contributions. We have:
\bea
{\cal L}_4&=&-\frac{1}{4}F_{\mu\nu}F^{\mu\nu}+\overline{l_L}i\gamma^\mu D_\mu l_L+\overline{l_R}i\gamma^\mu \partial_\mu l_R+\overline{q_L}i\gamma^\mu D_\mu q_L+\overline{q_R}i\gamma^\mu \partial_\mu q_R\nn\\
&&+D_\mu\varphi^\dagger D^\mu\varphi-V(\varphi^\dagger\varphi)-\left(y_l~\varphi \overline{l_L}l_R+y_q~\varphi^\dagger \overline{q_L}q_R+h.c.\right)~~~,
\eea
where $D_\mu \psi=(\partial_\mu+i g Q(\psi)A_\mu)\psi$, $(\psi=l_L,l_R,q_L,q_R,\varphi)$ and $V(\varphi^\dagger\varphi)=\mu^2(\varphi^\dagger\varphi)+\lambda(\varphi^\dagger\varphi)^2$ 
$(\lambda>0)$. Depending on the sign of $\mu^2$, the gauge symmetry is spontaneously broken or not. When $\mu^2<0(\mu^2>0)$
the theory is in the broken(unbroken) phase. We will discuss both cases at once.  At dimension six we have \cite{Grzadkowski:2010es}:
\be
{\cal L}_6=\sum_k c_k O_k+...
\ee
where $O_k$ are the operators in table 1, dots stand for four-fermion operators, that will be discussed in Section
\ref{4f4}.
The coefficients $c_k$ have mass dimension $-2$ and implicitly carry the dependence on some reference scale $\Lambda$~\footnote{It is customary to set: 
$c_k=\tilde c_k/\Lambda^2$, where $\tilde c_k$ are dimensionless.}.
We define the classical current $j^\mu$ as:
\be
j^\mu(x)=-\frac{1}{g}\frac{\delta S}{\delta A_\mu(x)}~~~.
\ee
Here $S$ is the classical action and the derivative is the variational one. We have:
\be
j^\mu=j^\mu_3+j^\mu_5~~~,
\label{current1}
\ee
\bea
j^\mu_3&=&\partial_\lambda F^{\lambda \mu}-\overline{l_L}\gamma^\mu l_L+\overline{q_L}\gamma^\mu q_L-i(\varphi^\dagger \Dfb^\mu\varphi)~~~,\nn\\
j^\mu_5&=&
-ic_{\varphi D}~\varphi^\dagger\varphi(\varphi^\dagger \Dfb^\mu\varphi)
+4 g c_{\varphi A}\partial_\lambda(\varphi^\dagger\varphi F^{\lambda\mu})
+4 g c_{\varphi \widetilde A}\partial_\lambda(\varphi^\dagger\varphi \widetilde F^{\lambda\mu})\nn\\
&&
+2 c_{l A}\partial_\lambda(\varphi \overline{l_L}\sigma^{\lambda\mu}l_R)
+2 c_{q A}\partial_\lambda(\varphi^\dagger \overline{q_L}\sigma^{\lambda\mu}q_R)
-2\varphi^\dagger\varphi\sum_i c_{\varphi f_i} \bar f_i \gamma^\mu f_i~~~.
\label{current2}
\eea
\begin{table}[t] 
\centering
\begin{tabular}{|c|c|}
\hline
$O_\varphi$&$(\varphi^\dagger\varphi)^3$\rule[-2ex]{0pt}{5ex}\\
\hline
$O_{\varphi\Box}$&$(\varphi^\dagger\varphi)\Box(\varphi^\dagger\varphi)$\rule[-2ex]{0pt}{5ex}\\
\hline
$O_{\varphi D}$&$(\varphi^\dagger D^\mu \varphi)^*(\varphi^\dagger D_\mu\varphi)$\rule[-2ex]{0pt}{5ex}\\
\hline
$O_{\varphi A}$&$g^2(\varphi^\dagger\varphi)F_{\mu\nu}F^{\mu\nu}$\rule[-2ex]{0pt}{5ex}\\
\hline
$O_{\varphi \widetilde A}$&$g^2(\varphi^\dagger\varphi)F_{\mu\nu}\widetilde F^{\mu\nu}$\rule[-2ex]{0pt}{5ex}\\
\hline
\end{tabular}~~~~~~~~~
\begin{tabular}{|c|c|}
\hline
$O_{l\varphi}$&$(\varphi^\dagger\varphi)\varphi \overline{l_L}l_R $\rule[-2ex]{0pt}{5ex}\\
\hline
$O_{q\varphi}$&$(\varphi^\dagger\varphi)\varphi^\dagger \overline{q_L}q_R $\rule[-2ex]{0pt}{5ex}\\
\hline
$O_{l A}$&$g\varphi \overline{l_L}\sigma^{\mu\nu}l_R F_{\mu\nu}$\rule[-2ex]{0pt}{5ex}\\
\hline
$O_{q A}$&$g\varphi^\dagger \overline{q_L}\sigma^{\mu\nu}q_R F_{\mu\nu}$\rule[-2ex]{0pt}{5ex}\\
\hline
$O_{\varphi l_L}$&$i(\varphi^\dagger \Dfb_\mu\varphi)\overline{l_L}\gamma^\mu l_L$\rule[-2ex]{0pt}{5ex}\\
\hline
$O_{\varphi l_R}$&$i(\varphi^\dagger \Dfb_\mu\varphi)\overline{l_R}\gamma^\mu l_R$\rule[-2ex]{0pt}{5ex}\\
\hline
$O_{\varphi q_L}$&$i(\varphi^\dagger \Dfb_\mu\varphi)\overline{q_L}\gamma^\mu q_L$\rule[-2ex]{0pt}{5ex}\\
\hline
$O_{\varphi q_R}$&$i(\varphi^\dagger \Dfb_\mu\varphi)\overline{q_R}\gamma^\mu q_R$\rule[-2ex]{0pt}{5ex}\\
\hline

\end{tabular}
\caption{Dimension-six operators other than the four-fermion ones.\label{tab:no4ferm}}
\end{table}
From the invariance of the classical action under infinitesimal gauge transformations $\delta_\alpha A_\mu(x)$ and
$\delta_\alpha \chi_I(x)$, $\chi_I(x)$ denoting collectively the matter fields, we get:
\be
\delta_\alpha S=\int d^4x \left[\frac{\delta S}{\delta A_\mu(x)}\delta_\alpha A_\mu(x)+
\frac{\delta S}{\delta \chi_I(x)}\delta_\alpha \chi_I(x)\right]=0~~~,
\ee
Along the solutions of the equations of motion, where the second term vanishes, we recover the
conservation of the current $j_\mu(x)$ (\ref{current1},\ref{current2}) in the classical theory.

We analyze the anomalous contributions caused by fermion loops
by means of the effective action $W[A_\mu,\varphi,\varphi^\dagger]$ defined in eq. (\ref{effact}). The infinitesimal variation of a generic functional $F[A_\mu,\varphi,\varphi^\dagger]$ under gauge transformations
of the fields $A_\mu$, $\varphi$ and $\varphi^\dagger$ can be expressed trough the operator $L(x)$:
\be
\delta_\alpha F=\int d^4 x \alpha(x)L(x)F~~~~~~~L(x)=\left[-\frac{1}{g}\partial_\mu\frac{\delta}{\delta A_\mu(x)}+i\varphi(x)\frac{\delta}{\delta \varphi(x)}-i\varphi^\dagger(x)\frac{\delta }{\delta \varphi^\dagger(x)}
\right]~.
\ee
From the invariance of the action $S$
under gauge transformations, the following identity follows:
\be
L(x)W[A_\mu,\varphi,\varphi^\dagger]={\tt Anomaly}(x)~~~,
\label{ward1}
\ee
where the anomaly is a local polynomial in the bosonic fields $B=(A_\mu,\varphi,\varphi^\dagger)$, which vanishes
provided the measure is invariant under a gauge transformation of the fermionic sector~\footnote{An equivalent, more familiar, statement is $\partial_\mu \langle j^\mu(x)\rangle_B+i\varphi(x)\frac{\delta W}{\delta \varphi(x)}-i\varphi(x)^\dagger\frac{\delta W}{\delta \varphi(x)^\dagger}={\tt Anomaly}$,
where, by definition, $\langle j^\mu(x)\rangle_B\equiv-\frac{1}{g}\frac{\delta W}{\delta A_\mu(x)}$.}.

This is precisely the issue analyzed here. When ${\cal L}_6$ is set to zero, the condition for anomaly cancellation, $Q(l_L)+Q(q_L)=0$, is automatically satisfied and eq. (\ref{ward1}) holds with vanishing right-hand-side. We would like to check whether the cancellation of the gauge anomaly requires additional conditions on the coefficients $c_k$, when ${\cal L}_6$ is turned on.  To this purpose we directly compute the functional $W$, controlled by
the interaction involving fermion bilinear terms:
\be
{\cal L}_{\bar f f}=\bar f \left(\mathscr{S}+\mathscr{P} \gamma_5+\mathscr{V}_\mu \gamma^\mu+\mathscr{A}_\mu \gamma^\mu\gamma_5+\mathscr{T}_{\mu\nu} \sigma^{\mu\nu} \right)f~~~,
\label{bilinear}
\ee
where $f=(l,q)^T$ and:
\bea
\mathscr{S}&=&\frac{1}{2}\left(\mathscr{Y}+\mathscr{Y}^\dagger \right)~~~~~~~~~~
\mathscr{P}=\frac{1}{2}\left(\mathscr{Y}-\mathscr{Y}^\dagger \right)\nn\\
\mathscr{V}_\mu&=&\frac{1}{2}\left[-gQA_\mu+i(C_{\phi R}+C_{\phi L})(\varphi^\dagger \Dfb_\mu\varphi)\right]\nn\\
\mathscr{A}_\mu&=&\frac{1}{2}\left[+gQA_\mu+i(C_{\phi R}-C_{\phi L})(\varphi^\dagger \Dfb_\mu\varphi)\right]\nn\\
\mathscr{T}_{\mu\nu}&=&\frac{g}{2}\left[(\Sigma+\Sigma^\dagger)F_{\mu\nu}+i(\Sigma-\Sigma^\dagger)\widetilde F_{\mu\nu}\right]~~~,
\eea
The quantities  $Q$, $C_{\phi R}$, $C_{\phi L}$ are field-independent matrices, whereas $\mathscr{Y}$, and $\Sigma$ are matrices depending on the scalar field $\varphi$ and its conjugate:
{\footnotesize
\bea
Q&=&
\left(
\begin{array}{cc}
-1&0\\
0&+1
\end{array}
\right)~~~~~~~
C_{\phi R(L)}=
\left(
\begin{array}{cc}
c_{\varphi l_{L(R)}}&0\\
0&c_{\varphi q_{L(R)}}
\end{array}
\right)\nn\\
\mathscr{Y}&=&
\left(
\begin{array}{cc}
-y_l\varphi+c_{l\varphi}\varphi^\dagger\varphi\varphi&0\\
0&-y_q\varphi+c_{q\varphi}\varphi^\dagger\varphi\varphi
\end{array}
\right)~~~~~~~
\Sigma=
\left(
\begin{array}{cc}
c_{lA}\varphi&0\\
0&c_{qA}\varphi^\dagger
\end{array}
\right)~~~.\nn
\eea}

\noindent
The functional $W$ depends on $A_\mu$, $\varphi$, $\varphi^\dagger$ through
the combinations $\mathscr{S}$, $\mathscr{P}$, $\mathscr{V}_{\mu}$, $\mathscr{A}_{\mu}$ and $\mathscr{T}_{\mu\nu}$.
To check the gauge invariance of $W$, we Taylor expand $W$ in powers of these combinations.
Since only triangle diagrams with $\mathscr{V}_{\mu}$ and $\mathscr{A}_{\mu}$ insertions can contribute to the anomaly \cite{Bardeen:1969md,Clark:1983ev,Bardeen:1985hr},
we only need to consider the following term:
\bea
W&=&\int d^4x_1 d^4x_2 d^4x_3
\left\{\frac{1}{3!} W^{LLL}_{\mu\nu\lambda}(x_1,x_2,x_3){\tt tr}[\mathscr{L}_{\mu}(x_1) \mathscr{L}_{\nu}(x_2) \mathscr{L}_{\lambda}(x_3)]\right.\nn\\
&&~~~~~~~~~~~~~~~~~~~+\frac{1}{2!} W^{LLR}_{\mu\nu\lambda}(x_1,x_2,x_3){\tt tr}[\mathscr{L}_{\mu}(x_1) \mathscr{L}_{\nu}(x_2) \mathscr{R}_{\lambda}(x_3)]\nn\\
&&~~~~~~~~~~~~~~~~~~~~+\frac{1}{2!} W^{LRR}_{\mu\nu\lambda}(x_1,x_2,x_3){\tt tr}[\mathscr{L}_{\mu}(x_1) \mathscr{R}_{\nu}(x_2) \mathscr{R}_{\lambda}(x_3)]\nn\\
&&~~~~~~~~~~~~~~~~~~~~+\left.\frac{1}{3!} W^{RRR}_{\mu\nu\lambda}(x_1,x_2,x_3){\tt tr}[\mathscr{R}_{\mu}(x_1) \mathscr{R}_{\nu}(x_2) \mathscr{R}_{\lambda}(x_3)]
\right\}+...
\eea
where we have defined:
\be
\mathscr{R}_{\mu}(x)=\mathscr{V}_{\mu}(x)+\mathscr{A}_{\mu}(x)~~~,~~~~~~~~~
\mathscr{L}_{\mu}(x)=\mathscr{V}_{\mu}(x)-\mathscr{A}_{\mu}(x)~~~.
\ee
The coefficients $W^{XYZ}_{\mu\nu\lambda}(x_1,x_2,x_3)$ are field-independent functions of the space-time
points $x_{1,2,3}$, that can be evaluated by a one-loop computation. The function
$W^{LLL}_{\mu\nu\lambda}(x_1,x_2,x_3)$ is symmetric under permutations of $(\mu,x_1)$, $(\nu,x_2)$ and 
$(\lambda,x_3)$. Analogous properties hold for the other expressions. 
We also define $W^{YXZ}_{\nu\mu\lambda}(x_2,x_1,x_3)=W^{XYZ}_{\mu\nu\lambda}(x_1,x_2,x_3)$ and so on.
An important property of the operator $L(x)$ is:
\be
L(x)\mathscr{L}_\mu(y)=Q~\partial_\mu \delta^4(x-y)~~~,~~~~~~~~~~~L(x)\mathscr{R}_\mu(y)=0~~~.
\label{lprop}
\ee
This is a consequence of the gauge invariance of the combination $i(\varphi^\dagger \Dfb_\mu\varphi)$ contributing to $\mathscr{V}_\mu$ and $\mathscr{A}_\mu$. We get:
\bea
L(x)W&=&\frac{1}{2}\int d^4 y d^4 z
\left\{\partial^\mu W^{LLL}_{\mu\nu\lambda}(x,y,z)~{\tt tr}[Q~\mathscr{L}_{\nu}(y)\mathscr{L}_{\lambda}(z)]
\right.\nn\\
&&~~~~~~~~~+2~\partial^\mu W^{LLR}_{\mu\nu\lambda}(x,y,z)~{\tt tr}[Q~\mathscr{L}_{\nu}(y)\mathscr{R}_{\lambda}(z)]\nn\\
&&~~~~~~~~~~~+\left.\partial^\mu W^{LRR}_{\mu\nu\lambda}(x,y,z)~{\tt tr}[Q~\mathscr{R}_{\nu}(y)\mathscr{R}_{\lambda}(z)]\right\}~~~.
\label{ward2}
\eea
We see that, as a consequence of eq. (\ref{lprop}), when evaluating the triangle diagrams we do not need to
insert the whole current $j_\mu(x)$ (\ref{current1}) (in the vertex which is acted upon by the derivative) but only the lowest dimensional part $j^\mu_3(x)$  \cite{Bonnefoy:2020tyv}. This is not yet sufficient to prove the independence of $L(x)W$ on the parameters $c_k$ of ${\cal L}_6$, since they still appear
in the combinations $\mathscr{L}_\mu$ and $\mathscr{R}_\mu$ at the other two vertices of the triangle.

To evaluate the correlators $\partial^\mu W^{XYZ}_{\mu\nu\lambda}(x,y,z)$ of eq. (\ref{ward2}) we first regularize the
effective action $W$ by using dimensional regularization \cite{tHooft:1972tcz,Marinucci:1975hx,Breitenlohner:1977hr} with the t'Hooft-Veltman prescription for $\gamma_5$
~\footnote{That is: $\{\gamma_5,\gamma^{\bar\mu}\}=0$ and $[\gamma_5,\gamma^{\hat\mu}]=0$, $\gamma^{\bar\mu}$ ($\gamma^{\hat\mu}$)
denoting the four-dimensional ($(d-4)$-dimensional) part of $\gamma^\mu$. 
Note that such prescription is at the origin of chirality-mixing contribution:
terms like $(1-\gamma_5)\slashed{k} \gamma^{\hat\mu} (1+\gamma_5)$ do not vanish 
and give rise to evanescent terms of order $(d-4)$. Due to the pole $1/(d-4)$
arising from the integration, such terms are converted into finite, chirally-mixed terms.}. In momentum space the correlators
are easily computed through well-known triangle diagrams. Denoting by $W_{\tt r}$ the regulated effective action,
we get:
\bea
L(x)W_{\tt r}=-\frac{1}{24\pi^2}\varepsilon^{\mu\nu\rho\sigma}
&&\Big\{\xi_{AA}~\partial_\mu A_{\nu}(x)\cdot \partial_\rho A_{\sigma}(x)\nn\\
&&+i~\xi_{\varphi A}~\partial_\mu (\varphi^\dagger \Dfb_\nu\varphi)(x)\cdot \partial_\rho A_{\sigma}(x)\nn\\
&&-\xi_{\varphi\varphi}~\partial_\mu (\varphi^\dagger \Dfb_\nu\varphi)(x)\cdot \partial_\rho (\varphi^\dagger \Dfb_\sigma\varphi)(x)\Big\}~~~,
\label{wouldbe}
\eea
where:
\bea
\xi_{AA}&=&g^2~{\tt tr}~Q^3\nn\\
\xi_{\varphi A}&=&-g~{\tt tr}~Q^2(2 C_{\phi L}+C_{\phi R})\nn\\
\xi_{\varphi\varphi}&=&{\tt tr}~Q(C_{\phi L}^2+C_{\phi L}C_{\phi R}+C_{\phi R}^2)~~~.\nn
\eea
As a consequence, the gauge variation of $W_{\tt r}$ is given by:
\be
\delta_\alpha W_{\tt r}=\int d^4 x \alpha(x)L(x)W_{\tt r}~~~,
\ee
and we see that it is still dependent on the coefficients $c_k$ through the combinations $\xi_{\varphi A}$ and $\xi_{\varphi\varphi}$. However,
the right-hand side of this expression should not be identified with the anomaly, since the effective action $W$
and the functional $W_{\tt r}$ differ by counterterms, space-time integrals of local polynomials in the bosonic fields, that can be added
to $W_{\tt r}$.  
If a counterterm $W_{\tt c}$ exists such that $\delta_\alpha W_{\tt c}+\delta_\alpha W_{\tt r}=0$, we can define
$W=W_{\tt r}+W_{\tt c}$ and the theory is free from gauge anomalies. In general, we expect $L(x)W_{\tt r}$ to contain relevant and irrelevant contributions. This is the case in the model under consideration. 
The first line in eq. (\ref{wouldbe}) is relevant. A local counterterm whose gauge variation cancels this term does not exist.
Gauge anomaly cancellation requires the condition ${\tt tr}Q^3=0$, satisfied by construction in this model.
The remaining part of $L(x)W_{\tt r}$ is irrelevant. Indeed consider:
\bea
W_{\tt c}=\frac{1}{24\pi^2}\varepsilon^{\mu\nu\rho\sigma}\int d^4 y
&&\Big\{+i~\xi_{\varphi A}~g A_\mu(y) (\varphi^\dagger \Dfb_\nu\varphi)(y)\cdot \partial_\rho A_{\sigma}(y)\nn\\
&&-\xi_{\varphi\varphi}~g A_\mu(y) (\varphi^\dagger \Dfb_\nu\varphi)(y)\cdot \partial_\rho (\varphi^\dagger \Dfb_\sigma\varphi)(y)\Big\}~~~.
\eea
We have:
\bea
L(x)W_{\tt c}=-\frac{1}{24\pi^2}\varepsilon^{\mu\nu\rho\sigma}\int d^4y
&&\Big\{+i~\xi_{\varphi A}~\partial_\mu\delta^4(x-y) (\varphi^\dagger \Dfb_\nu\varphi)(y)\cdot \partial_\rho A_{\sigma}(y)\nn\\
&&-\xi_{\varphi\varphi}~\partial_\mu\delta^4(x-y)(\varphi^\dagger \Dfb_\nu\varphi)(y)\cdot \partial_\rho (\varphi^\dagger \Dfb_\sigma\varphi)(y)\Big\}\nn\\
=
\frac{1}{24\pi^2}\varepsilon^{\mu\nu\rho\sigma}
&&\Big\{+i~\xi_{\varphi A}~\partial_\mu (\varphi^\dagger \Dfb_\nu\varphi)(x)\cdot \partial_\rho A_{\sigma}(x)\nn\\
&&-\xi_{\varphi\varphi}~\partial_\mu(\varphi^\dagger \Dfb_\nu\varphi)(x)\cdot \partial_\rho (\varphi^\dagger \Dfb_\sigma\varphi)(x)\Big\}~~~.\nn\\
\label{counter}
\eea
We see that $L(x)W_{\tt c}$ cancels the second and third lines in eq. (\ref{wouldbe}). Hence, by choosing $W=W_{\tt r}+W_{\tt c}$
we end up with:
\be
L(x) W=-\frac{g^2{\tt tr}Q^3}{24\pi^2}\varepsilon^{\mu\nu\rho\sigma}~\partial_\mu A_{\nu}(x)\cdot \partial_\rho A_{\sigma}(x)~~~~.
\label{end}
\ee
Thus the condition for gauge anomaly cancellation has no dependence on the coefficients of the higher
dimensional operators defining the EFT. It is completely controlled by the charge assignment $Q$ of the
fermion sector. Some comments are in order.
\begin{itemize}
\item[$\bullet$]
The gauge invariance of the combination $i(\varphi^\dagger D_\nu\varphi-D_\mu \varphi^\dagger \varphi)$ is a crucial ingredient for the result (\ref{end}).
\item[$\bullet$]
Non-trivial gauge-invariant local polynomials like those of eq. (\ref{cand}), which would have contributed to the anomaly,
have not shown up in $L(x)W_{\tt r}$.
\item[$\bullet$]
The result (\ref{end}) holds whether
the gauge symmetry is realized in the unbroken phase or not. If the theory is in the broken phase, the whole
combination $i(\varphi^\dagger D_\nu\varphi-D_\mu \varphi^\dagger \varphi)$ should be included in $L(x)W_{\tt r}$.
Had we kept only the first term of this expression in the expansion around the vacuum $\varphi=v/\sqrt{2}$, that is $i(\varphi^\dagger D_\nu\varphi-D_\mu \varphi^\dagger \varphi)=v^2 g A_\mu+...$, the second and third lines in eq. (\ref{wouldbe})
would have collapsed to expressions proportional to the first line and we would have missed the cancellation displayed above.
\item[$\bullet$]
The $c_k$-dependent terms shown in eqs. (\ref{wouldbe}) and (\ref{counter}) might depend on the regularization used.
If we compute $L(x)W_{\tt r}$ using a set of Pauli-Villars regulators \cite{Bilal:2008qx}, we might get
different coefficients $\xi_{\varphi A}$ and $\xi_{\varphi\varphi}$. In this case, the same cancellation mechanism is at work,
but with a different choice of counterterms. This reflects the unphysical nature of the $c_k$-dependent part.
Only the sum of the regulated diagrams and counterterms is a physical quantity. This has a direct impact on
the amplitude for a physical process, evaluated beyond the tree-level approximation.
If we do not include the contribution of the counterterm $W_{\tt c}$ to the set of relevant diagrams, 
in general we will get a gauge-dependent result. Hence the counterterms highlighted above
have not a purely academic interest, but play an essential role in the evaluation of a physical quantity.
\end{itemize}
To complete the discussion, we still need to include the contribution from four-fermion operators, which we do in the next Section.

\subsection{Four-fermion operators}
\label{4f4}
Four-fermion operators can be easily accounted for, without spoiling our conclusion.
Consider a complete set of four-fermion operators. Here again we focus on the $B$ and $L$ conserving sector. Through Fierz transformations we can cast their
contribution to the Lagrangian in the form:
\be
{\cal L}_6^{4F}=\frac{1}{2}~\bar l \Gamma_I l~ C^{ll}_{IJ}~ \bar l \Gamma_J l+\frac{1}{2}~\bar q \Gamma_I q~ C^{qq}_{IJ}~ \bar q \Gamma_J q+\bar l \Gamma_I l~ C^{lq}_{IJ}~ \bar q \Gamma_J q~~~,
\label{fourf}
\ee
where $\Gamma_I$ represents the set $(1,\gamma_5,\gamma^\mu,\gamma^\mu\gamma_5,\sigma^{\mu\nu})$
and $C^{ll,qq,ql}_{IJ}$ are matrices of coefficients. We can equivalently express the combination in (\ref{fourf})
in terms of fermion bilinears, by making use of a set of bosonic auxiliary fields $\chi_I=(\mathnormal{s},\mathnormal{p},\mathnormal{v}^\mu,\mathnormal{a}^\mu,\mathnormal{t}^{\mu\nu})$ with suitable masses and couplings \cite{Soto:1990ij}:
\be
{\cal L}_4^{\tt aux}=\chi_I~ \bar f
\left(
\begin{array}{cc}
X^l_{IJ}&0\\
0&X^q_{IJ}
\end{array}
\right) \Gamma_J f-\frac{1}{2}\chi_I M^2_{I}\chi_I~~~.
\ee
We can always choose $X^{l,q}_{IJ}$ and $M_I^2$ such that, by eliminating the auxiliary fields $\chi_I$ through their equations of motion, we reproduce eq. (\ref{fourf}). Notice that the gauge invariance of ${\cal L}_6^{4F}$ implies
the gauge invariance of all the auxiliary fields $\chi_I$ in ${\cal L}_4^{\tt aux}$. The new term ${\cal L}_4^{\tt aux}$
modifies the bilinear fermion interaction of eq. (\ref{bilinear}) by adding to the combinations $\mathscr{S}$, $\mathscr{P}$, $\mathscr{V}_{\mu}$, $\mathscr{A}_{\mu}$ and $\mathscr{T}_{\mu\nu}$ new gauge invariant contributions. In particular,
the quantities $\mathscr{V}_{\mu}$ and $\mathscr{A}_{\mu}$, relevant to the computation of $L(x)W$ are now modified into:
\bea
\mathscr{V}_\mu&=&\frac{1}{2}\left[-gQA_\mu+i(C_{\phi R}+C_{\phi L})(\varphi^\dagger \Dfb_\mu\varphi)+C_V\mathnormal{v}_\mu\right]\nn\\
\mathscr{A}_\mu&=&\frac{1}{2}\left[+gQA_\mu+i(C_{\phi R}-C_{\phi L})(\varphi^\dagger \Dfb_\mu\varphi)+C_A\mathnormal{a}_\mu\right]~~~,
\eea
where $C_{V,A}$ are matrices easily identifiable from the couplings $X^{l,q}_{IJ}$. The effective action $W$ now depends
on the bosonic background through the gauge field $A_\mu$ and the gauge invariant quantities $i(\varphi^\dagger D_\nu\varphi-D_\mu \varphi^\dagger \varphi)$, $\mathnormal{v}_\mu$ and $\mathnormal{a}_\mu$. The gauge variation $L(x)W_{\tt r}$
is a linear combination of:
\be
\varepsilon^{\mu\nu\rho\sigma}\partial_\mu B^a_{\nu}(x)\cdot \partial_\rho B^b_{\sigma}(x)~~~,
\ee
where $B^a_\mu=(A_\mu,i(\varphi^\dagger D_\nu\varphi-D_\mu \varphi^\dagger \varphi),\mathnormal{v}_\mu,\mathnormal{a}_\mu)$. Only the term with $B^a_\mu=B^b_\mu=A_\mu$ is non-trivial. By following the same steps
of the previous section, we see that the rest is irrelevant and we conclude that the presence of four-fermion interactions does not alter the condition of gauge anomaly cancellation.

\section{Discussion}
There are several arguments supporting the idea that the cancellation of gauge anomalies 
in a general EFT only depends on the set of fermion representations and not
on the features of the non-renormalizable sector. Nevertheless, an explicit computation of the gauge anomaly
in a non-trivial EFT, including the full set of allowed operators up to a given dimensionality and allowing
for the most general bosonic background is still missing. This note was meant to fill this gap and to elucidate
the mechanism that removes the dependence of the anomaly on the higher dimensional operators. 
An additional motivation is provided by the ineffectiveness of cohomological arguments
to rule out contributions to the anomaly of arbitrarily high dimension.
Indeed, when the gauge group is non semi-simple, candidate anomalies are local gauge-invariant polynomials 
whose dimension is in principle unbounded.

In this note we have carried out an explicit computation of the anomaly in a simple abelian model, with
several features in common with the SMEFT. We have included the most general set of dimension six operators
comprising also four-fermion operators. We have evaluated the anomaly within dimensional regularization
both for diagrams involving minimal fermion interactions and for those where fermions are non-minimally coupled.
After the inclusion of appropriate counterterms, the resulting anomaly is independent on the coefficients of 
the non-renormalizable sector. Counterterms are expected to depend on the adopted regularization and only
the sum of all diagrams, counterterms included, has a physical meaning. Thus the counterterms
highlighted here have a direct impact when amplitudes for physical processes are evaluated beyond the tree-level approximation
within dimensional regularization.

We can identify two crucial ingredients in our derivation.
Firstly, in the presence of the higher-dimensional operators, the gauge current acquires a new gauge invariant term,
that does not contribute to the gauge variation of the effective action. This point has been recently stressed in ref. 
\cite{Bonnefoy:2020tyv}.
Secondly, the variation of the effective action evaluated in a general bosonic background includes relevant and irrelevant terms. Only the irrelevant component depends on the coefficients of the non-renormalizable operators. This component can be subtracted by adding a local counterterm to the Lagrangian density. 
We expect that also in the SMEFT a similar mechanism takes place. Indeed we do not foresee a qualitatively different 
behaviour of the SMEFT under a generic gauge transformation. Nevertheless, we think that it is important to fully
identify the set of counterterms needed to cancel the spurious non-invariance induced by higher-dimensional SMEFT
operators. Indeed, counterterms of this type have already been adopted  in the 
automation of one-loop computations in the SMEFT, on a case by case basis \cite{Durieux:2018ggn,Degrande:2020evl}.
The general computation illustrated in this note might allow to determine the whole set of such
counterterms, covering all possible processes involving operators of dimension six.

\section*{Acknowledgements}
I am grateful to Claudia Cornella for reporting me an error in expression of the coefficients
$\xi_{\varphi A}$ and $\xi_{\varphi\varphi}$, in the first version of this work.
I thank Cen Zhang for signaling to me the works of ref. \cite{Durieux:2018ggn,Degrande:2020evl}. 
This project has received support by INFN and by the European Union's Horizon 2020 research and innovation programme under the Marie Sklodowska-Curie grant agreement N$^\circ$~860881-HIDDeN.
\newpage


\end{document}